\documentclass[aps,prl,twocolumn,floatfix]{revtex4}
\usepackage{graphicx}
\graphicspath{{Figures/}}

\def\gtwid{\mathrel{\raise.3ex\hbox{$>$\kern-.75em\lower1ex\hbox{$\sim$}}}}
\def\alt{\mathrel{\raise.3ex\hbox{$<$\kern-.75em\lower1ex\hbox{$\sim$}}}}

\def\agt{\mathrel{\raise.3ex\hbox{$>$\kern-.75em\lower1ex\hbox{$\sim$}}}}
\def\alt{\mathrel{\raise.3ex\hbox{$<$\kern-.75em\lower1ex\hbox{$\sim$}}}}

\newcommand{\be}{\begin{equation}}
\newcommand{\ee}{\end{equation}}

\begin{document}

\title{Reynolds numbers of the large-scale flow in turbulent Rayleigh-B\'enard convection} 
\author{Guenter Ahlers, Eric Brown, Denis Funfschilling, and Alexei Nikolaenko}
\address{Department of Physics and iQCD, University of California, Santa Barbara, California 93106}
\date{\today}

\begin{abstract}
We measured Reynolds numbers $R_e$ of turbulent Rayleigh-B\'enard convection over the Rayleigh-number range $2\times 10^8 \alt R \alt 10^{11}$ and Prandtl-number range $3.3 \alt \sigma \alt 29$ for cylindrical samples of aspect ratio $\Gamma = 1$. For $R \alt R_c \simeq 3\times 10^9$ we found $R_e \sim R^{\beta_{eff}}$ with $\beta_{eff} \simeq 0.46 < 1/2$. Here both the $\sigma$- and $R$-dependences are  quantitatively consistent with the Grossmann-Lohse (GL) prediction. For $R > R_c$ we found $R_e = 0.106~ \sigma^{-3/4} R^{1/2}$, which differs from the GL prediction. The relatively sharp transition at $R_c$ to the large-$R$ regime suggests a qualitative and sudden change that renders the GL prediction inapplicable.
\end{abstract}


\maketitle

Understanding turbulent Rayleigh-B\'enard convection (RBC) in a fluid heated from below \cite{Si94} remains one  of the challenging problems in nonlinear physics. It is well established that a major component of the dynamics of this system is a large-scale circulation (LSC) \cite{KH81}. For cylindrical samples of aspect ratio $\Gamma \equiv D / L \simeq 1$ ($D$ is the diameter and $L$ the height) the LSC consists of a single convection roll, with both down-flow and up-flow near the side wall but at azimuthal locations $\theta$ that differ by $\pi$. An additional important component of the dynamics is the generation of localized volumes of relatively hot or cold fluid, known as ``plumes", at a bottom or top thermal boundary layer. The hot (cold)  plumes are carried by the LSC from the bottom (top) to the top (bottom) of the sample and by virtue of their buoyancy contribute to the maintenance of the LSC. The LSC plays an important role in many natural phenomena, including atmospheric and oceanic convection, and convection  in the outer core of the Earth where it is believed to be responsible for the generation of the magnetic field.  In this Letter we report measurements of the speed of the LSC that agree well with a theoretical prediction by Grossmann and Lohse \cite{GL02} for relatively small applied temperature differences $\Delta T$, but depart from this prediction rather suddenly as $\Delta T$ is increased further.  Our results illustrate clearly that a quantitative understanding of this system is still restricted to limited parameter ranges.

The LSC can be characterized by a turnover time $\cal T$ and an associated Reynolds number \cite{QT02}
\begin{equation}
R_e = (2L/{\cal T})\times(L/ \nu)
\label{eq:Re}
\end{equation}
($\nu$ is the kinematic viscosity). A central prediction of various theoretical models \cite{Si94,Kr62, GL00,GL01,GL02,GL04} is the dependence of $R_e(R,\sigma)$ on the Rayleigh number 
\begin{equation}
R = \alpha g \Delta T L^3/\kappa \nu
\label{eq:R}
\end{equation}
and on the Prandtl number 
\begin{equation}
\sigma = \nu/\kappa
\end{equation}
($\alpha$ is the isobaric thermal expansion coefficient, $\kappa$ the thermal diffusivity, and $g$ the acceleration of gravity). A recent prediction by Grossmann and Lohse (GL) \cite{GL02}, based on the decomposition of the 
kinetic and the thermal dissipation into boundary-layer and bulk contributions,  has been in remarkably good agreement with experimental results for $R_e(R,\sigma)$ \cite{ReFN}. However, the parameter range covered by the measurements was relatively small.

We report new measurements of $Re(R,\sigma)$ over a wider range,  for $R$ up to $10^{11}$ and $3.3 \alt \sigma \alt 29$. For modest $R$, say $R \alt 2\times 10^9$, we again find very good agreement with the predictions of GL. However, for larger $R$ the measurements reveal a relatively sudden transition to a new state of the system, with a Reynolds number that is described well by
\be
R_e = 0.106~\sigma^{-3/4}~R^{1/2}\ .
\label{eq:Re_exp}
\ee
This result differs both in the $\sigma$ dependence and in the $R$ dependence from the GL prediction. We interpret our results to indicate the existence of a new LSC state. It is unclear at present whether the difference between this state and the one at smaller $R$ will be found in the geometry of the flow, in the nature of the viscous boundary layers that  interact with it, or in the nature and frequency of plume shedding by the thermal boundary layers adjacent to the top and bottom plates. But whatever its nature, this state does not conform to the consequences of the assumptions made in the GL model.

Another important aspect of the predictions is the dependence of the Nusselt number (the dimensionless effective thermal conductivity)
\begin{equation}
{\cal N} = Q L / \lambda \Delta T
\label{eq:nusselt}
\end{equation}
on  $R$ and $\sigma$ (here  $Q$ is the heat-current density and $\lambda$ the thermal conductivity). 
The GL model \cite{GL01,GL02} provides a good fit also to data for ${\cal N}$ at modest $R$, say up to $R \simeq 10^{10}$ \cite{AX01,XLZ02,FBNA05}. Here we briefly mention as well  measurements of $\cal N$ for larger $R$ \cite{FBNA05} that depart significantly from the GL prediction as $R$ approaches $10^{11}$. 

Measurements of $R_e$ were made for three cylindrical samples with $\Gamma \simeq 1$. Two of them, known as the medium and large sample, \cite{BNFA05} had $L = 24.76$ and 49.69 cm respectively. The third was similar to the small sample of Ref. \cite{BNFA05}, but had $L = 9.52$. As evident from Eq.~\ref{eq:R}, a given accessible range of $\Delta T$ will provide data over different ranges of $R$ for the different $L$ values. For the small sample we used 2-propanol with $\sigma = 28.9$ as the fluid and measured the frequency $f$ of oscillations of the direction of motion of plumes across the bottom plate to obtain  $R_e = 2L^2f/\nu$ \cite{FA04}. With the medium and large sample we used water, mostly at mean temperatures $T_m = 55.00,~40.00$, and 29.00$^\circ$C  corresponding to $\sigma = 3.32,~4.38,~5.55$ 
and $\nu = 5.11\times 10^{-7},~6.69\times 10^{-7},~8.25\times 10^{-7}$ m$^2/$sec respectively. 
The top and bottom plates were made of copper. A plexiglas side wall had a thickness of 0.32 (0.63) cm for the medium (large) sample. At the horizontal  mid-plane  eight thermistors, equally spaced around the circumference and labeled $i = 0, ..., 7$, were imbedded in small holes drilled horizontally into but not penetrating the side wall. The thermistors were able to sense the adjacent fluid temperature without interfering with delicate fluid-flow structures. When a given thermistor (say $i=0$) sensed a relatively high temperature $T_i$ due to warm upflow of the LSC, then the one located on the opposite side (say at $i = 4$) would sense a relatively low temperature due to the relatively cold downflow.

\begin{figure}
 \includegraphics[width=3in]{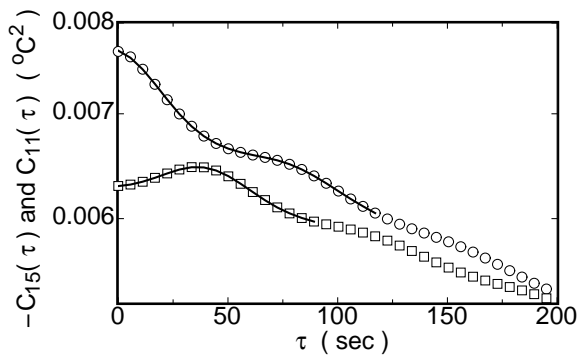}
\caption{The cross-correlation function (squares) between two thermometers (1 and 5) mounted on opposite sides of the side wall, and the auto-correlation function (circles) of a single thermometer (1), on a logarithmic scale as a function of the delay time $\tau$ for the large sample and $R = 7.5\times 10^{10}$. Solid lines: fits of Eqs.~\ref{eq:acfit} or \ref{eq:ccfit} to the data. The lengths of the lines indicate the range of the data used in the fits.}
\label{fig:cor}
\end{figure}

When a warm (cold) plume passed a given side-wall thermistor, the indicated temperature  was relatively high (low). It had been shown before \cite{QT02}, by comparison of temperature sensors actually imbedded in the fluid and laser-doppler velocimetry, that this thermal signature can be used to determine the speed, and thus the Reynolds number, of the LSC and that it yields the same result as actual velocity measurements. Indeed, where there is overlap, our results for $R_e$ are in satisfactory agreement with measurements \cite{QT02} based on velocimetry.

From time series of the eight temperatures $T_i(t)$ taken at intervals of a few seconds and covering at least one and in some cases more than ten days at each of many values of $R$ we determined the auto-correlation functions (AC) $C_{i,j}(\tau),~i=j$, and the cross-correlation functions (CC) $C_{i,j}(\tau),~ i = 0, \ldots, 3,~ j = i+4$ corresponding to signals at azimuthal positions displaced around the circle by $\pi$. They are given by 

\begin{equation}
C_{i,j}(\tau) = \langle [T_i(t) - \langle T_i(t)\rangle_t]\times[T_j(t+\tau) -  \langle T_j(t)\rangle_t] \rangle_t\ .
\label{eq:corfunc}
\end{equation}

\noindent We show an example of AC (circles) and of CC (squares) in Fig.~\ref{fig:cor}. 

One sees that the AC have a peak centered at the origin. It can be represented well by a Gaussian function. The peak width  indicates that the plume signal is correlated over a significant time interval. A second smaller Gaussian peak is observed at a later time $t_2^{ac}$ that we identify with one turn-over time ${\cal T}$ of the LSC. The existence of this peak indicates that the plume signal retains some coherence while the LSC undergoes a complete rotation \cite{Vi95}. A further very faint peak is found at $2{\cal T}$, but is not used in our analysis. These observations are consistent with previous experiments \cite{CGHKLTWZZ89,TBM93,QT02}. This structure is superimposed onto a broad background that decays roughly exponentially on a time scale of ${\cal O}(10{\cal T})$. We believe that the background decay is caused by a slow meandering of the azimuthal orientation of the LSC.

The CC are consistent with the AC. Here too there is a broad, roughly exponential, background. There is no peak at the origin, and  the first peak, of Gaussian shape, occurs at a time delay $t_1^{cc} = {\cal T}/2$ associated with half a rotation of the LSC. A further peak is observed at $3{\cal T}/2$, corresponding to 1.5 full rotations.

Based on the above, we fitted the equation 

\begin{eqnarray}
C_{i,i}(\tau) &=& b_0 exp\left (- \frac{\tau}{\tau^{ac}_0}\right ) + b_1 exp \left [- \left (\frac{\tau}{\tau^{ac}_1} \right )^2 \right ]\nonumber \\
 &+& b_2 exp \left [-\left ( \frac{\tau - t^{ac}_2}{\tau^{ac}_2}\right )^2 \right ]
\label{eq:acfit}
\end{eqnarray}

\noindent to the data for the AC, and the equation

\begin{equation}
C_{i,j}(\tau) = - b_0 exp \left ( -\frac{\tau}{\tau^{cc}_0} \right ) - b_1 exp \left [ -\left ( \frac{\tau - t^{cc}_1}{\tau^{cc}_1}\right )^2 \right]
\label{eq:ccfit}
\end{equation}

\noindent to those for the CC. Examples of the fits are shown in Fig.~\ref{fig:cor} as solid lines. One sees that the fits are excellent.

\begin{figure}
 \includegraphics[width=3.25in]{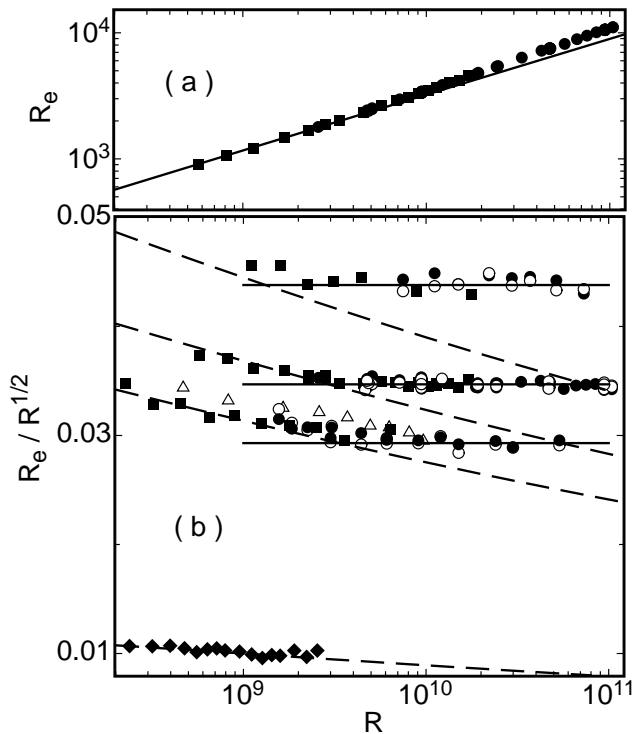}
\caption{(a): Reynolds numbers $R_e^{cc}$ for $\sigma = 4.38$ as a function of the Rayleigh number $R$ on logarithmic scales. (b): $R_e/R^{1/2}$ on a linear scale, as a function of $R$ on logarithmic scales. From top to bottom, the four data sets are for $\sigma = 3.32$, 4.38, 5.55, and 28.93. Solid (open) circles: $R^{cc}_{e}$ ($R^{ac}_{e}$), large sample. Solid squares: $R^{cc}_{e}$, medium sample. Solid diamonds: $R_e^\omega$ for 2-propanol at 40$^\circ$C. Open triangles: Ref. \cite{QT02}, $\sigma = 5.4$. 
Dashed lines (from top to bottom): GL predictions for $\sigma = 3.32, ~4.38, ~5.55,$ and 28.9. \cite{FN} Horizontal solid lines: $R_e/R^{1/2} = 0.0433,$ 0.0347, and 0.0293.}
\label{fig:Recc}
\end{figure}

Substituting ${\cal T} = t^{ac}_2$ and  ${\cal T} = 2 t^{cc}_1$ into Eq.~\ref{eq:Re}, we have 
\begin{equation}
R^{ac}_{e,i} = 2 (L^2/\nu) / t^{ac}_2\ .
\label{eq:Reac}
\end{equation}
and
\begin{equation}
R^{cc}_{e,i} =  (L^2/\nu) / t^{cc}_1\ .
\label{eq:Recc}
\end{equation}
as two experimental estimates of $R_e$.
For each $R$ we computed the average value $R_e^{cc}$ of the eight CC $C_{i,j}$ and $C_{j,i}$ with $i = 0,  \ldots, 3$ and $j = i+4$. The results are shown in Fig.~\ref{fig:Recc} as solid squares (medium sample) and solid circles (large sample). Averages of the eight AC $C_{i,i}$ at each $R$ for the large sample are shown as open circles.  There is excellent agreement between the AC and the CC.  Also shown, as solid diamonds, are results for the small sample deduced from the oscillation of the direction of plume motion across the bottom plate \cite{FA04}. These data are for 2-propanol with $\sigma = 28.9$.
For comparison, the results of Qiu and Tong \cite{QT02,FN_QT} based on velocity measurements for $\sigma = 5.4$ are shown as open triangles. Our data for $\sigma = 5.55$ are in quite good agreement with them. 

The dashed lines in Fig.~\ref{fig:Recc} are, from top to bottom, the predictions of GL \cite{GL02,FN} for $\sigma = 3.25,~4.38,~5.55$ and 28.9. For $R \alt 3\times 10^9$ they pass very well through the data. We regard this agreement of the prediction with our measurements as a major success of the model. However, for larger $R$ the data quite suddenly depart from the prediction and scatter randomly about the horizontal solid lines. These results indicate that there is a sudden change of the exponent $\beta_{eff}$ of the power law
\be
R_e(R,\sigma) = R_0 \sigma^{-\alpha_{eff}} R^{\beta_{eff}}
\label{eq:eff_powerlaw}
\ee 
 as $R$ exceeds $R_c \simeq 2\times 10^9$, from a value less than 1/2 to 1/2 within experimental resolution. The GL model can not reproduce this behavior, and we conclude that a new large-$R$ state is entered that does not conform  to the assumptions made in the model.

\begin{figure}
 \includegraphics[width=2.5in]{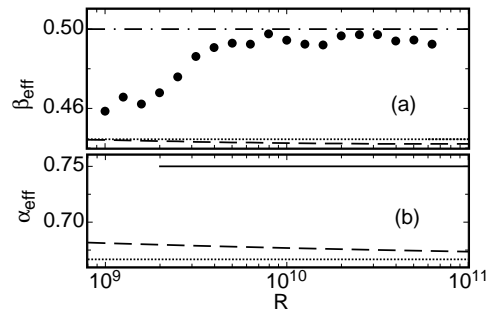}
\caption{Effective exponents $\beta_{eff}$ (a) and $\alpha_{eff}$ (b) of Eq.~\ref{eq:eff_powerlaw} as a function of $R$. Dotted lines: asymptotic values predicted by GL for $R \rightarrow \infty$. Dashed lines: effective values as a function of $R$ predicted by GL for $\sigma=4.38$. Solid circles in (a): experimental values for $\beta_{eff}$ obtained by fitting a powerlaw to the data for $R_e(R)$ at $\sigma = 4.38$, using a sliding window 0.8 decades wide. Dash-dotted line in (a): $\beta_{eff} = 1/2$. Solid line in (b): approximate location of experimental results.}
\label{fig:exponents}
\end{figure}

\begin{figure}
 \includegraphics[width=2.5in]{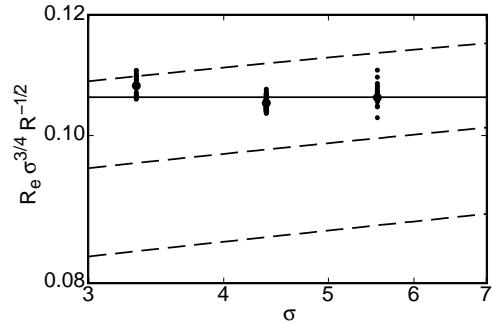}
\caption{The reduced Reynolds number $R_e \sigma^{3/4} R^{-1/2}$ as a function of the Prandtl number $\sigma$ on logarithmic scales. Dashed lines (from top to bottom): GL predictions for $R = 10^9,~10^{10},$ and $10^{11}$ \cite{FN}. Solid line: Eq.~\ref{eq:Re_exp}. Dots: all data for $3\times 10^9 \alt R \alt 10^{11}$.}
\label{fig:Re_of_sig}
\end{figure}

\begin{figure}
 \includegraphics[width=3in]{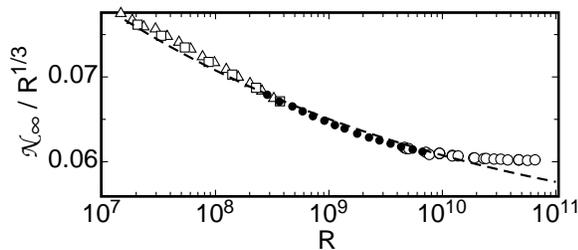}
\caption{The reduced Nussel number ${\cal N}_\infty/R^{1/3}$ as a function of $R$ for $\sigma = 4.38$ \cite{FBNA05}. Dashed line: GL prediction.}
\label{fig:nusselt}
\end{figure}

The inconsistency between the prediction and the data can bee seen more clearly by considering the effective exponents $\beta_{eff}$ and $\alpha_{eff}$ defined by Eq.~\ref{eq:eff_powerlaw} and shown in Fig.~\ref{fig:exponents}. The experimental values of $\beta_{eff}$ in Fig.~\ref{fig:exponents}a were obtained by fitting powerlaws to the data for $\sigma = 4.38$, using a sliding window 0.8 decades wide. One sees that, within experimental uncertainty, the value 1/2 is reached at $R \simeq 7\times 10^{9}$. As can be  seen from Fig.~\ref{fig:Recc}, $\beta_{eff} \simeq 1/2$ actually is reached earlier, near $R \simeq 3\times 10^9$;  the results in Fig.~\ref{fig:exponents}a represent an average over a finite range of $R$ because  a finite window width had to be employed in the analysis. The GL model predicts the value $\beta = 4/9\simeq 0.444$ when $R$ becomes large enough so that a pure power law prevails (dotted line). The  predicted effective values $\beta_{eff}$ at finite $R$ (dashed line) are already very close to this value in the experimental range of $R$. It is hard to see how the prediction could be changed by adjusting parameters in the model so as  to yield $\beta_{eff} = 1/2$ for $R \agt 2\times  10^9$ without changing the seemingly firm prediction $\beta_{eff} \rightarrow 4/9$ for sufficiently large $R$.

Figure~\ref{fig:exponents}b shows $\alpha_{eff}$ defined by Eq.~\ref{eq:eff_powerlaw}. Here the GL prediction yields $\alpha = 2/3$ as $R$ becomes large (dotted line). The  dashed line gives the predicted $\alpha_{eff}$ as a function of $R$ for finite $R$. One sees that $\alpha_{eff}$ is already quite close to $\alpha$ in the experimental range of $R$. In the range  where the experimental $R_e/R^{1/2}$ is constant (i.e. $2\times 10^9 \alt R \alt 10^{11}$) the data are consistent with $\alpha_{eff} = 3/4$ as shown by the  solid line, but not with $\alpha_{eff} \simeq 2/3$. To explore this point further, we show in Fig.~\ref{fig:Re_of_sig} $R_e\sigma^{3/4}/R^{1/2}$ as a function of $\sigma$. Here all our data for  $R \agt 3\times 10^9$ are plotted. We note that, at each $\sigma$,  all data  collapse into a narrow range  consistent with the scatter of the measurements. The horizontal solid line, which corresponds to Eq.~\ref{eq:Re_exp} and thus to $\alpha_{eff} = 3/4$,  passes through the points at each  $\sigma$ within the scatter, although it is a bit low for the smallest $\sigma$. The solid lines are the GL predictions for, from top to bottom,  $R = 10^9,~10^{10},$ and $10^{11}$.

It is interesting to note that a similar inconsistency was found also between the GL prediction and the Nusselt number \cite{FBNA05}. This is illustrated in Fig.~\ref{fig:nusselt} where we show the reduced Nusselt number ${\cal N}_{\infty}/R^{1/3}$ as a function of $R$. There are deviations from the prediction \cite{GL01}  (dashed line) for $R \agt 10^{10}$, which is somewhat higher than the value of $R_c$ for the Reynolds number. 

In this Letter we presented new measurements of the Reynolds number $R_e$ of the large-scale circulation in turbulent Rayleigh-B\'enard convection for an aspect-ratio-one cylindrical sample over the Rayleigh-number range $2\times 10^8 \alt R \alt 10^{11}$ and the Prandtl-number range $3.3 \alt \sigma \alt 29$. For $R \alt 3\times 10^9$, where $R_e \propto R^{0.46}$, our data agree well with the prediction by Grossmann and Lohse \cite{GL02}; but for larger $R$ we find that $R_e = 0.106 \sigma^{3/4} R^{1/2}$, in disagreement with the GL prediction.

We thank Xin-Liang Qiu for providing us with the numerical data corresponding to Fig. 12 of Ref. \cite{QT02}. This work was supported by the US Department of Energy through Grant  DE-FG02-03ER46080.


\begin{thebibliography}{3}
\bibitem{FN} We changed the parameters of the GL model to read $c_1 = 10.16, ~c_2 = 1.832, ~c_3 = 0.478, ~c_4 = 0.0141, ~a = 0.464, ~Re_c = 0.925$. This gives a better fit to our $R_e$ data and does not alter the prediction for ${\cal N}(R,\sigma)$ \cite{GL02}.
\end{thebibliography}
\end{document}